\def\year{2016}\relax
\documentclass[letterpaper]{article}
\pdfoutput=1
\usepackage{aaai16copyrightfix}
\usepackage{times}
\usepackage{graphicx}
\usepackage{helvet}
\usepackage{courier}
\begin{document}

\title{Qualitative Framing of Financial Incentives -- A Case of Emotion Annotation}
\author{Sephora Madjiheurem \and Valentina Sintsova \and Pearl Pu \\
\'Ecole Polytechnique F\'ed\'erale de Lausanne Switzerland \\ \{sephora.madjiheurem$|$valentina.sintsova$|$pearl.pu\}@epfl.ch }
\maketitle
\begin{abstract}
Online labor platforms, such as the Amazon Mechanical
Turk, provide an effective framework for eliciting responses
to judgment tasks. 
Previous work has shown that workers
respond best to financial incentives, especially to extra bonuses. 
However, most of the tested incentives 
involve describing the bonus conditions in formulas instead of plain English.
We believe that different incentives given in English (or in qualitative framing)
will result in differences in workers' performance, especially when
task difficulties vary.
In this paper, we report
the preliminary results of a crowdsourcing experiment comparing
workers' performance using only qualitative
framings of financial incentives.
Our results demonstrate a significant increase in workers' performance
using a specific well-formulated qualitative framing
inspired by the Peer Truth Serum. This positive effect is observed  only when the difficulty of
the task is high, while when the task is easy there is
no difference of which incentives to use.
\end{abstract}

\section{Introduction}
In online labor markets, the quality of the answers returned by inexpert, remote workers presents an important concern for the usefulness of such crowdsourcing process. The~standard techniques for quality assurance, such as combining answers from different workers and posterior removal of non-satisfactory answers \cite{ipeirotis2010quality}, assume the existence of a correct answer. 
Yet, for judgment tasks---tasks involving human subjective judgment such as collecting opinions, preferences, relevance estimations, ratings, or emotion categorization---it is difficult to establish the validity of a specific answer based only on the~(dis)agreement with the majority label. 

Previous research showed the potential of motivating workers by introducing additional incentives, such as bonuses for good-quality answers \cite{ho2015incentivizing,harris2011you}. Yet, these works require automatic estimation of the answers' correctness. 
The schemas without this requirement would be preferable for judgment tasks. 
In this direction, other works suggest that giving bonus based on the consistency with peer answers is most advantageous \cite{Shaw:2011:DII:1958824.1958865,huang2013enhancing,faltings2014d}. 
Such peer-oriented schemas originate from a game-theoretic approach to incentivize truthful answers \cite{prelec2004}, and thus they imply the mathematical formulation of bonus computation. However, a lay person from Amazon Mechanical Turk (MTurk) is unlikely to fully understand the implications of such computations, especially in case of non-mathematical judgment tasks (such as emotion labeling). 
To overcome this, we suggest to study alternative simplified formulations of the peer-oriented schemas suitable for application in judgment tasks without any mathematical formulas. We refer to this approach as qualitative framing. 
We report in this paper the results of a crowdsourcing experiment conducted on the MTurk, where we compare several qualitatively framed peer-oriented incentives with the ones inspired by previous research on social incentives in crowdsourcing \cite{Shaw:2011:DII:1958824.1958865} across two difficulty levels in the context of emotion annotation. 

\section{Experiment Design}

We use in our experiment a specific judgment task---annotating emotions in tweets. 
We ask each worker to label $10$ tweets after completing a mandatory tutorial. For each tweet, the worker should  indicate the dominant emotion felt by the author while writing the tweet by selecting one of $20$ emotion categories, \textit{No~emotion}, or \textit{Other~emotion} from the Geneva Emotion Wheel (GEW, version~2.0) \cite{scherer2013grid}.
 For example, for the tweet \textit{``Woooo! It's a good day"}, the worker is likely to select the emotion \textit{Happiness}.
The worker is also asked to provide the excerpts of the tweet indicating the presence of the chosen emotion (\textit{``Woooo"} and \textit{``good day"} in the example) as well as additional emotion indicators (not from the tweet text). The same task was presented and studied in our previous work \cite{sintsova-musat-pu:2013:WASSA}. We also included optional demographic and feedback questionnaires.
For the sake of brevity, we report here only the results for the emotion labels.

We randomly assigned each worker to an incentive treatment condition employing a specific bonus formulation. 
To discover whether the effects of incentives differ depending on the difficulty of the annotation, we assigned workers at random to one of two difficulty levels.
 
The base payment was set to $\$0.5$ USD, and there was information about a chance to obtain an additional bonus of $\$0.1$ USD. A specific description of the bonus (dependent on the assigned incentive condition) was presented to a worker right before starting the tweets' annotation:

\noindent\textit{``You can qualify for the additional $\$0.1$ bonus if your answers [...]" } 

\vspace{1pt}\noindent
\textbf{Normative: } 
\textit{``... demonstrate an additional effort." }

\vspace{1pt}\noindent
\textbf{Experts:}
\textit{``... are extremely accurate according to our experts." }

\vspace{1pt}\noindent
\textbf{Professors: }
\textit{``... are approved by our professors."}

\vspace{1pt}\noindent
\textbf{Peer Agreement: }
\textit{``... agree with those of other workers." }

\vspace{1pt}\noindent
\textbf{Peer Truth Serum 1 (PTS1): }
\textit{``... are more surprisingly common with other workers than collectively predicted."} \\
\textbf{Peer Truth Serum 2 (PTS2): }
\textit{``... both agree with those of other workers and at the same time novel."} 

\vspace{4pt}\noindent
The second variable was the difficulty of the tweets to label: 

\vspace{2pt}\noindent
\textbf{Easy dataset} The first, ``easy'' dataset consisted of $70$ tweets manually chosen such that they were explicitly emotional and not difficult to interpret, judged by us as requiring less effort to annotate.  
The example tweet is \textit{``You said it would be different but like usual nothing has changed"}.

\vspace{2pt}\noindent
\textbf{Difficult dataset} The second, ``difficult'' dataset comported $70$ emotional tweets that were less obvious to interpret and believed to require more attention to annotate, because they contained negated emotional terms, or were of a sarcastic nature, or simply confusing. The example tweets are \textit{``I do not regret a single thing being born as Dusun."} and  \textit{``Best part about rush hour is driving into it going Chicago!!! \#not"}.

\section{Results}
We ran the experiment on MTurk between May 13th and June 3rd  2015, with $1,135$ workers completing the full task. 
The random assignment of workers to treatment conditions resulted in a fairly even split of the non-spamming workers across all conditions ($94.6 \pm 3.2$ workers per group).

We estimate agreement and correctness of the returned emotion labels to evaluate the workers' performance. We report average per-worker scores. 
The computation of each metric for one worker's answer on a tweet is as follows.
\textit{Category agreement} of a worker's emotion category label is computed as the percentage of agreed peer labels for the same tweet. 
 To compute \textit{category correctness}, we first obtain the ground-truth categories by aggregating the workers' labels from all incentive conditions. 
  For each tweet, we extract the majority label and all the categories with a relatively high assignment number (we use the threshold of $0.5$ minimum ratio from the majority label count). The worker's label is considered to be correct if it is within the set of extracted ground-truth labels for the tweet.

\begin{figure}[t]
\centering
\includegraphics[width=\linewidth]{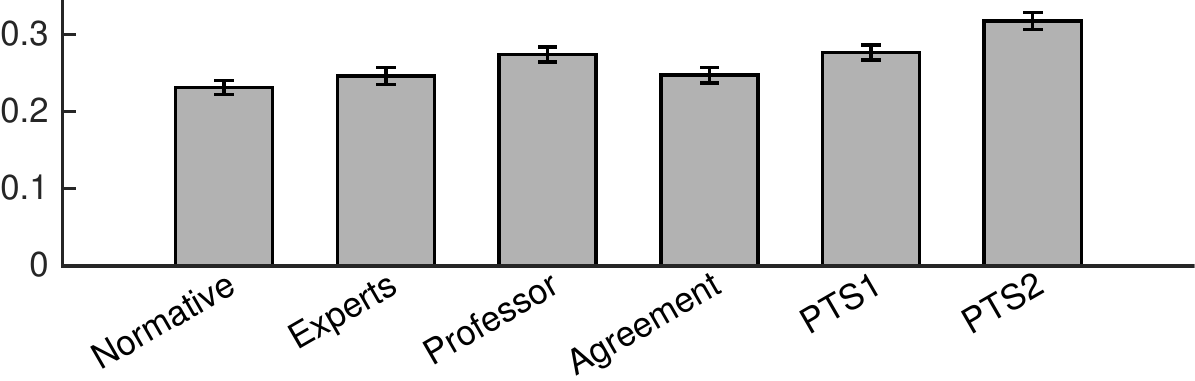}
\caption{The effect of incentives on average \textit{category agreement} of workers assigned to the difficult dataset. Error bars indicate a standard error.
}
\label{fig:diffCat}
\end{figure}

\vspace{2pt}
Because the two-way ANOVA revealed a significant interaction between the two studied factors, 
we decided to resort on analyzing the simple effects of incentives at each level of difficulty.
We found \textit{no significant effect  of the incentives' framing on the quality of emotion labels on the easy dataset} (p-value of one-way ANOVA is $0.183$ for category agreement and $0.571$ for category correctness on the easy dataset). 
Nevertheless, the incentives significantly affect worker's performance on the difficult dataset  
Indeed, p-value of one-way ANOVA is~$<10^{-7}$ for category agreement and $0.033$ for category correctness. 
Figure \ref{fig:diffCat} visualizes the differences  in category agreement between incentives' conditions on the difficult dataset.
Pairwise comparison of the incentives reveals the\textit{ advantage of PTS2 incentive on the difficult dataset}: it is significantly better than any other incentive in terms of category agreement (the highest p-value is $0.045$ when comparing with PTS1), and it leads to the highest category correctness while significantly outperforming the Normative condition (with p-value~$=0.023$). 
The Normative condition, which is appealing only to workers' honesty, in its turn results in the lowest category agreement on the difficult dataset, significantly different from PTS1, PTS2, and Professors incentives (with p-values~$\le0.034$). 

The workers' answers to the feedback questionnaire verify that the difficult dataset is indeed perceived as more difficult than the easy dataset and that it requires additional cognitive effort to label. 
This implicitly approves our methodology for selecting tweets for the difficult and easy datasets.
The further statistical analysis revealed no effect of the incentives' framing on the workers' perceived task comprehension, enjoyment, easiness, or cognitive effort.

\section{Discussion}

The results of our experiment revealed the differences in the effects of the incentives' framing depending on the difficulty of the emotion annotation task.  When the task is easy, no incentive results in significantly improved performance compared to the other incentives. However, when difficulty of the task increases, our findings suggest that it is beneficial to use a particular incentive: with a well-formulated version of the  Peer Truth Serum (PTS) bonus, workers tend to output more correct and agreeing emotion labels.
 This finding is in accordance with benefits of this schema in its mathematical formulation for non-judgment tasks \cite{faltings2014d}. We study two different qualitative formulations of the Peer Truth Serum (\textit{``surprisingly common"} vs. \textit{``agreeing and novel"}) and found that only the second formulation can lead to significantly better answers  showing the importance of better qualitative framing of the incentives. 

The fact that the qualitative framing of incentives affects the quality of answers at least for one fixed bonus amount encourages the future investigation, testing, and deployment of more advantageous qualitative bonus formulations.
Our  findings could be of potential value to the researchers and practitioners aiming to design incentives mechanisms for judgment tasks. 

\bibliography{incentives-short-paper.bib} 
\bibliographystyle{aaai}

\end{document}